# Exciton Superposition across Moiré States in a Semiconducting Moiré Superlattice

Zhen Lian[1#], Dongxue Chen[1#], Yuze Meng[1], Xiaotong Chen[1], Ying Su[2], Rounak Banerjee[3], Takashi Taniguchi[4], Kenji Watanabe[5], Sefaattin Tongay[3], Chuanwei Zhang[2], Yong-Tao Cui[6*], Su-Fei Shi[1,7*]

1. Department of Chemical and Biological Engineering, Rensselaer Polytechnic Institute, Troy, NY 12180, USA
2. Department of Physics, University of Texas, Dallas, Texas, 75083, USA
3. School for Engineering of Matter, Transport and Energy, Arizona State University, Tempe, AZ 85287, USA
4. International Center for Materials Nanoarchitectonics, National Institute for Materials Science, 1-1 Namiki, Tsukuba 305-0044, Japan
5. Research Center for Functional Materials, National Institute for Materials Science, 1-1 Namiki, Tsukuba 305-0044, Japan
6. Department of Physics and Astronomy, University of California, Riverside, California, 92521, USA
7. Department of Electrical, Computer & Systems Engineering, Rensselaer Polytechnic Institute, Troy, NY 12180, USA

[#] These authors contributed equally to this work
[*] Corresponding authors: shis2@rpi.edu, yongtao.cui@ucr.edu

## Abstract

**Moiré superlattices of semiconducting transition metal dichalcogenides (TMDCs) enable unprecedented spatial control of electron wavefunctions in an artificial lattice with periodicities more than ten times larger than that of atomic crystals, leading to emerging quantum states with fascinating electronic and optical properties[1,2]. The breaking of translational symmetry further introduces a new degree of freedom inside each moiré unit cell: high symmetry points of energy minima called moiré sites, behaving as spatially separated quantum dots[3–5]. The superposition of a quasiparticle's wavefunction between different moiré sites will enable a new platform for quantum information processing but is hindered by the suppressed electron tunneling between moiré sites. Here we demonstrate the superposition between two moiré sites by constructing an angle-aligned trilayer $WSe_2$/monolayer $WS_2$ moiré heterojunction. The two moiré sites with energy minimum allow the formation of two different interlayer excitons, with the hole residing in either moiré site of the first $WSe_2$ layer interfacing the $WS_2$ layer and the electron in the third $WSe_2$ layer. An external electric field can drive the hybridization of either of the interlayer excitons with the intralayer excitons in the third $WSe_2$ layer, realizing the continuous tuning of interlayer exciton hopping between two moiré sites. Therefore, a superposition of the two interlayer excitons localized at different moiré sites can be realized, which can be resolved in the electric-field-dependent optical reflectance spectra, distinctly different from that of the natural trilayer $WSe_2$ in which the moiré modulation is absent. Our study illustrates a strategy of harnessing the new moiré site degree of freedom for quantum information science, a new direction of twistronics.**

**Main Text**

Design and control of symmetries can lead to symmetry-protected states that are promising to revolutionize the field of quantum materials. For example, breaking the inversion symmetry or time-reversal symmetry can lead to Weyl semimetals[6]. The inversion symmetry breaking in transition metal dichalcogenides (TMDCs) gives rise to a valley degree of freedom that is promising for valleytronics and quantum information science based on valley-spin[7], which can be accessed through chiral light.

The recent emergence of semiconducting TMDC moiré superlattices[1–3,8–12], which are constructed through twisted TMDCs with a lattice mismatch or twist angle, enables spatial control of the excitons in two-dimension (2D) with the tunable periodicity of 1-10 nm and ushers in unprecedented opportunities in engineering electrons and excitons, leading to intriguing correlated electronic states[1,8,13–18], the array of quantum emitters, and correlated exciton states resulting from flat excitonic band[19–21].

Translation symmetry breaking in TMDC moiré superlattices introduces a new degree of freedom: moiré sites, the high symmetry points in a moiré supercell, which can be local energy minima and act as quantum dots that can confine electrons and excitons[4], as schematically shown in Fig. 1c. In addition, these high symmetry points are protected by the three-fold rotation symmetry and possess the unique valley degree of freedom through the pseudo angular momentum conservation[2,4]. As a result, coupling and hybridization of these high symmetry points will usher in new venues toward quantum information storage and processing. However, unlike the energy degeneracy of different valleys (K and K'), the energy barrier between different moiré sites (on the order of 10s' meV), along with their spatial separation, greatly suppresses the direct coupling between these moiré sites.

Here, we demonstrate the superposition between two different moiré sites by introducing a layer degree of freedom to the TMDC moiré superlattice. It is well known that the two neighboring layers of 2-H TMDC flakes, due to the intralayer inversion symmetry breaking, possess a layer degree of freedom that acts as pseudospins alternating in odd and even layers[7,22]. In an angle-aligned trilayer $WSe_2$/monolayer $WS_2$ heterojunction (3L $WSe_2$/ 1L $WS_2$), new types of interlayer excitons emerge, with holes residing in the first $WSe_2$ layer either trapped in moiré A or B site (Fig. 1c)[3], and electrons with the same pseudospin residing in the third $WSe_2$ layer. In particular, we find that these two interlayer excitons can hybridize through coupling with the intralayer excitons in the third $WSe_2$ layer. The resulting hybridized exciton inherits both the large oscillator strength from the intralayer excitons and the sensitive electric field dependence from the moiré interlayer excitons[9,23–25]. More interestingly, by applying an electric field, we can drive the transition between the two interlayer moiré excitons' hybridization with the intralayer exciton in the third $WSe_2$ layer, enabling the continuous tuning of hopping of the interlayer exciton from 100% at one moiré site to 100% at the other, which is otherwise suppressed. In between the transition points, we obtain an excitonic complex that is the superimposition of the interlayer excitons that are otherwise localized at moiré A and B sites.

The schematic of the 3L WSe$_2$/ 1L WS$_2$ moiré heterojunction is shown in Fig. 1a, which is fabricated into a dual-gated device structure in which the doping and electric field can be independently controlled. We also fabricated a device of a dual-gated 2-H phase trilayer WSe$_2$ (3L WSe$_2$) (schematically shown in Fig. 1b) for the control study.

The doping-dependent optical reflectance contrast spectra of the 3L WSe$_2$/ 1L WS$_2$ heterojunction device are shown in Fig. 2e, which is evidently different from that of the natural trilayer (3L) WSe$_2$ device (Fig. 2f). The most pronounced resonance for the natural trilayer WSe$_2$ (Fig. 2f) is the intralayer exciton resonance X$_A$, which is at ~ 1.70 eV at zero doping, redshifted compared to the A exciton resonance in monolayer WSe$_2$ (~ 1.73 eV)[12]. X$_A$ is redshifted linearly for both n and p doping in a symmetric fashion, with a slope of ~1.3 meV/10$^{12}$cm$^{-2}$. $IX_{3L}$ are the interlayer excitons with the hole and electron separated in the first and third WSe$_2$ layer, which have two degenerate modes as schematically shown in Fig. 2c and are named as $IX_{3L}^+$ and $IX_{3L}^-$ ("+" and "-" denote the direction of the dipole moment in the sample coordinate. The direction of the positive electric field or dipole moment is defined as from top gate to back gate in 3L WSe$_2$, and from WSe$_2$ to WS$_2$ in 3L WSe$_2$/ 1L WS$_2$.). $IX_{3L}^{2s}$ is the 2s state of the $IX_{3L}$. The natures of $IX_{3L}$ and $IX_{3L}^{2s}$ become obvious in our later discussion of the electric field dependent reflectance contrast spectra (Fig. 3). Zoom-in of Fig. 2f with enhanced contrast is plotted in Extended Fig. 2 to show $IX_{3L}$ and $IX_{3L}^{2s}$ more clearly. Accompanying X$_A$ is a less pronounced resonance X$_{A'}$ with a larger slope (2.7 meV/10$^{12}$cm$^{-2}$). X$_{A'}$ is likely the exciton resonance of the middle (second) layer WSe$_2$ and is not the focus of this work (see detailed discussion in Supplementary Section 4).

In the optical reflectance contrast spectra of 3L WSe$_2$/ 1L WS$_2$ moiré heterojunction (Fig. 2e), $X_M^I$ is the previously discovered moiré intralayer exciton peak in the first layer WSe$_2$ interfacing WS$_2$, with the exciton trapped at the moiré A site. The doping dependence of $X_M^I$ clearly show the signature of the correlated insulating states at the filling factor of 1 and -1, corresponding to one electron and one hole per moiré superlattice, which was discussed in our previous publication[26]. On the p-doping side, the exciton resonances of X$_A$ and X$_{A'}$ are labeled as such due to their similar behaviors compared with that from the trilayer WSe$_2$ (Fig. 2f), with a redshift slope of 1.0 and 2.1 meV/10$^{12}$cm$^{-2}$ respectively. The n-doping side is different because the electrostatically introduced electrons are in the WS$_2$ layer instead of the WSe$_2$ layers due to the type II alignment, leaving the WSe$_2$ layers charge-neutral. We identify the X$_A$ and X$_{A'}$ in the n-doping side through their slopes as well, 1.0 and 2.1 meV/10$^{12}$cm$^{-2}$, respectively, the same as those in the p-doping region. The abrupt blueshift of the X$_A$ in the n-doping side (starts at n>1 and resonant energy around 1.725 eV) is likely due to the built-in electric field on WSe$_2$ layers arising from the electron accumulation in WS$_2$. We leave the related discussion in Supplementary Information Section 5. The focus of our work here is on the interlayer excitons within the 3L WSe$_2$ of the 3L WSe$_2$/ 1L WS$_2$ moiré heterojunction, with their schematics shown in Figs. 2a, b. The $IX_{3L}^+$ branch is visible and pronounced at the blue arrow in Fig. 2e, partially because it hybridizes with intralayer excitons and gains some oscillator strength but also because it retains the extended nature of interlayer excitons, hence sensing dielectric environment change associated with the Mott insulator transition at filling of one electron per moiré superlattice (n=1). The natures of both resonances are revealed in our later discussion of the electric field dependence study (Fig. 4).

The electric field-dependent reflectance contrast spectra of the trilayer WSe$_2$ device is shown in Fig. 3a, which is symmetric about the electric field due to its symmetric structure. The most noticeable feature is the "cross" pattern originating from the electric field evolution from interlayer exciton $IX_{3L}$. The slopes of each branch of the cross are roughly the same. These are arising from the Stark shift of the interlayer exciton $IX_{3L}$, with the two degenerate modes ($IX_{3L}^-$ and $IX_{3L}^+$) shifting oppositely under an electric field due to the dipole moment of opposite polarity. The Stark energy shift can be expressed as $\Delta E = -edF$, where F is the local electric field, e is the electron charge, and d is the electron and hole separation. We extract the value of d to be about 1.26 nm for both $IX_{3L}^-$ and $IX_{3L}^+$, which is about twice that of interlayer exciton dipole moment in WSe$_2$/WS$_2$ (0.7 nm)[9], confirming that the electron and hole of interlayer exciton occupy the two outside WSe$_2$ layers in a natural trilayer WSe$_2$.

The level avoiding at the intralayer exciton A (~ 1.70 eV) in Fig. 3a arises from the hybridization of the interlayer exciton and intralayer exciton. In the 2H trilayer WSe$_2$, there is significant tunneling of holes between the first and third layer WSe$_2$ as they have the same valley-layer pseudo spin, allowing the hybridization of the interlayer excitons with the intralayer excitons in either the first or third WSe$_2$ layer[27], as schematically shown in Fig. 3d. This hybridization can be well captured by a coupled two-level system, which is given by the following Hamiltonian in the basis of intralayer exciton and interlayer exciton:

$$\begin{bmatrix} X_a & \Delta \\ \Delta & X_i(F) \end{bmatrix}$$

where $X_a$ is the energy of the intralayer exciton, $X_i(F)$ is the energy of the interlayer exciton at a given electric field F, $\Delta$ is the coupling strength (see Supplementary Information Section 3 for details).

Take the positive electric field (direction defined in Fig. 3d) scenario as an example (Fig. 3c): a linearly dispersed interlayer exciton $IX_{3L}^+$ (white dotted line in Fig. 3c) and a non-dispersed intralayer exciton X$_A$ (black dotted line) can be used to well fit the observed hybridized spectra (red and blue dashed lines). From the fitting, we extract the coupling strength to be 10.7 ± 0.3 meV, larger than the linewidth of the hybridized exciton (~ 9.0±0.3 meV). The scenario of the negative electric field is similar, where the other interlayer exciton mode, $IX_{3L}^-$, hybridizes with the intralayer exciton (X$_A$) when the energy of the two excitons is tuned to resonance via the electric field. It is worth noting that we ignore the conduction band hybridization of the first and third layer WSe$_2$, which is theoretically predicted to be nonzero but orders of magnitude smaller than the holes[27]. The neglection of the conduction band hybridization is also justified by the electric-field-dependent reflectance contrast spectra of 3L WSe$_2$/ 1L WS$_2$, which is asymmetric about positive and negative electric fields (later discussion of Fig. 4).

The additional level avoiding at the energy around 1.79 eV in Fig. 3a is due to the hybridization of the interlayer exciton ($IX_{3L}$) with the 2s state of intralayer A exciton (Fig.3a and Extended Figs. 4a,b). The second level avoiding at higher energy (~1.80 eV) is due to the hybridization of the excited state of the interlayer exciton ($IX_{3L}^{2s}$) and 2s of the A exciton, which we enhance the contrast and show in Extended Figs. 4a, b. It is interesting to note that the energy difference between the ground state and 2s of interlayer exciton $IX_{3L}$ is about 51 meV, smaller but at the same order of magnitude compared with the energy difference between 2s and 1s of A exciton for trilayer WSe$_2$ (~95 meV, Extended Figs. 4a,b), suggesting the strongly bound nature of the interlayer exciton $IX_{3L}$. All these hybridization features are absent in a dual-gated nature bilayer WSe$_2$ (Extended Fig. 8), which is AB stacked with two layers of different layer pseudospin, further confirming our interpretation. The electric-field-dependent reflectance contrast spectra of a 4L WSe$_2$ device (Extended Fig. 7) show similar hybridization features but with two "crosses" slightly shifted in energy, about 10 meV. According to the interpretation of the 3L WSe$_2$ data, these two crosses are the two types of interlayer excitons from the 1st and 3rd layer WSe$_2$ and the 2nd and 4th layer WSe$_2$, which slightly shift in energy due to possible dielectric environment differences[28,29].

We now turn to the study of the electric field-dependent reflectance contrast spectra of the 3L WSe$_2$/ 1L WS$_2$ moiré heterojunction, shown in Fig. 4c. The negative electric field side has some similarity compared with that from trilayer WSe$_2$, while the positive electric field side is significantly different. More specifically, the hybridized spectrum on the positive electric field side involves three exciton branches: two dispersive (grey and white dotted lines in Fig. 4e) and one non-dispersive (black dotted line in Fig. 4e) branch. The necessity of involving three exciton branches is also obvious from the derivative of Fig 4e with respect to the electric field, as shown in Extended Fig. 3d. The two dispersive excitons have a similar slope for the Stark shift, translating to electron and hole separations of 1.313 ± 0.004 nm and 1.609 ± 0.004 nm (fitting details in Supplementary Information Section 3). Therefore, they are the interlayer excitons, similar to $IX_{3L}^+$ in the natural trilayer WSe$_2$, with the hole in the first WSe$_2$ layer interfacing WS$_2$ and the electron in the third WSe$_2$ layer away from the interface. The two different interlayer excitons stem from the moiré coupling modified valence band of the first WSe$_2$ layer. As schematically shown in Figs. 4a, the moiré modulation folds the valence band of the first WSe$_2$ layer into moiré minibands. The two interlayer excitons correspond to holes occupying the two moiré minibands located at different moiré sites[5], which effectively behave as two spatially separated quantum dots. Each of them is located at an energy minimum at a high symmetry point within the moiré unit cell, which we call moiré A and B sites, respectively. We thus label these two interlayer excitons as $IX_{3L}^{+\,(A)}$ and $IX_{3L}^{+\,(B)}$. Since the WS$_2$ and first WSe$_2$ layer are aligned at 60 degrees (H stacked) as determined by the second harmonic generation (SHG) spectra (Extended Fig. 5), the moiré A and B sites correspond to the $H_h^h$ and $H_h^X$ stacking configurations shown in Fig. 4b. The energy separation of the two interlayer excitons at zero electric field, 69 meV, represents the energy difference between the top two moiré minibands, if we ignore the difference in exciton binding energy.

This value is consistent with the energy difference between the intralayer excitons trapped at moiré A and B sites in the WSe$_2$/WS$_2$ moiré superlattice, ~53 meV[26]. The remaining non-dispersive branch corresponds to the intralayer exciton, $X_A$, with both hole and electron in the third WSe$_2$ layer. Therefore, $IX_{3L}^{+\,(A)}$, $IX_{3L}^{+\,(B)}$, and $X_A$ hybridize by sharing the electron in the third WSe$_2$ layer. The above picture of hybridization involving two moire interlayer excitons are confirmed by a control device (D4) of 3L WSe$_2$/ 1L WS$_2$ in the dual-gate configuration, with an intentionally misaligned angle (20-degree) between WSe$_2$ and WS$_2$ layers. The electric-field-dependent reflectance contrast spectra (Extended Fig. 9) indeed become symmetric about the electric field and similar to that of natural 3L WSe$_2$, and they show no signs of interlayer moiré excitons ($IX_{3L}^{+\,(A)}$ and $IX_{3L}^{+\,(B)}$).

It is worth noting that direct tunneling between moiré A and B sites is suppressed due to the energy barrier, their spatial separation, and different stacking symmetry. Therefore, a direct hybridization between these two sites is difficult to achieve, unlike the degenerate valley-spin bands in TMDCs. However, with the assistance from the mobile intralayer exciton in the third WSe$_2$ layer, hybridization of moiré A and B sites is realized, and we can controllably tune the interlayer excitons $IX_{3L}^{+}$ between moiré A and B sites. In fact, the hybridized exciton notated with the cyan dashed line is a mixture of interlayer excitons localized at the moiré A site and B site, with the probability tunable from 100% at A to 100% at B site by controlling the electric field (Fig. 4f).

On the negative electric field side, the interlayer exciton involved in the hybridization is $IX_{3L}^{-}$, with the hole in the third layer WSe$_2$ not experiencing the moiré modulation. Meanwhile, the intralayer exciton in the 1$^{st}$ WSe$_2$ layer is modified by the moiré potential to have a lower energy of ~ 1.667 eV and is trapped at the moiré A site, which is labeled as $X_M^I$. As a result, hybridization occurs between $IX_{3L}^{-}$ and $X_M^I$. Their coupling strength is extracted to be 11.4 ± 0.1 meV. The interlayer exciton $IX_{3L}^{-}$ can also couple to the other moiré excitons from the 1$^{st}$ layer WSe$_2$, which contributes to the weak features in Fig. 4 and is shown with enhanced contrast in Extended Figs. 4c,d.

The asymmetry of Fig. 4c between the n- and p-doping sides further justifies our neglection of conduction band hybridization: $IX_{3L}^{-}$ near hybridization region d in Fig. 4c goes directly through $X_A$, and $IX_{3L}^{+}$ near region e goes directly through $X_M^I$, with neither showing level avoiding. If the conduction band hybridization is significant, we should observe the hybridization of two interlayer excitons (due to moiré modulated conduction bands) and $X_A$. Similarly, the interlayer exciton from region e ($IX_{3L}^{+}$) in Fig. 4c should hybridize with $X_M^I$. We include a detailed discussion in Supplementary Information Section 6.

In summary, we have demonstrated a strategy to realize continuous tuning of interlayer exciton hopping between different moiré sites in the 3L WSe$_2$/ 1L WS$_2$ moiré superlattice. These additional degrees of freedom enable the formation of a tunable honeycomb lattice of excitons with exciting opportunities for engineering new quantum states. For example, considering the large spin-orbit coupling in TMDCs, the continuous tuning of the hopping can be potentially exploited for constructing Dirac and Weyl modes of excitons, as well as the topologically protected edge states connecting these modes[4]. Our demonstration of the superposition of excitons across the different moiré sites also inspires new venues of quantum information processing and harnessing the new moiré site degree of freedom for twistronics.

## Methods:

### Sample Fabrication

We used the same dry pick-up method[30] as reported in our earlier work to fabricate TMDC heterostructures[20,26]. The gold electrodes are pre-patterned on the Si/SiO$_2$ substrate. The monolayer TMDC flakes, BN flakes, and few-layer graphene (FLG) flakes are exfoliated on silicon chips with 285 nm thermal oxide. The thickness of BN flakes was determined by atomic force microscopy (AFM). The layer numbers of WSe$_2$ flakes were identified by optical contrast with the assistance of second-harmonic generation (SHG). Top BN and bottom flakes with equal thickness were intentionally used for devices D1, D2, and D3. The polycarbonate (PC)/ polydimethylsiloxane (PDMS) stamp was used to pick up TMDC monolayer and other flakes sequentially. The alignment of each layer is achieved under a home-built microscope transfer stage with the rotation controlled with an accuracy of 0.02 degrees. The PC is then removed in the chloroform/isopropanol sequence and dried with nitrogen gas. The final constructed devices were annealed in a vacuum (<10$^{-6}$ torr) at 250 °C for 8 hours.

### Optical measurements

During the optical measurements, a home-built confocal imaging system was used to focus the laser onto the sample (with a beam spot diameter ~ 2 µm) and collect the optical signal into a spectrometer (Princeton Instruments). The reflectance contrast measurement was performed using a supercontinuum laser source (YSL photonics). A relative flat reflectance background $R_0$ was obtained by fitting the reflectance spectrum at high hole-doping level with a polynomial function for each measured spot (see SI for details). The reflectance contrast is defined as $\frac{dR}{R} = \frac{R - R_0}{R_0}$. The reflectance contrast from device D1 and D2 are added by 0.3 and -0.3 for better presentation in the log scale. All optical spectroscopy measurements were performed at the temperature of 10 K with a Montana cryostat. The polarized SHG measurements were performed with a pulsed laser excitation centered at 900 nm (Ti: Sapphire; Coherent Chameleon) with a repetition rate of 80 MHz and a power of 80 mW. The crystal axes of the sample were fixed. A half-waveplate was placed between the beam splitter and the objective and was rotated to change the polarization angles of both the excitation laser and the SHG signal.

### Doping electric field calculations

The density of carriers introduced by the electrostatic gating is given by $n_e(n_p) = C_{tg}(V_{tg} - V_{tg}^0) + C_{bg}(V_{bg} - V_{bg}^0)$, where $C_{tg}(C_{bg})$ are the geometry capacitance of the top gate (back gate) and $V_{tg}(V_{bg})$ are the top gate (back gate) voltage. $V_{tg}^0$ and $V_{bg}^0$ are the onset gate voltages of the top gate and the back gate, determined experimentally from the regions where the 2s peaks remain visible. The electrical field in the TMDC is given by $F = \varepsilon_{BN}/\varepsilon_{TMDC}(V_{tg}/2d_1 - V_{bg}/2d_2)$, where $d_1$ ($d_2$) is the thickness of the top (bottom) BN determined by atomic force microscopy, $\varepsilon_{BN} = 3.5$ and $\varepsilon_{TMDC} = 7.2$ are the relative dielectric constants of h-BN and TMDC, respectively[31,32].

## Data Availability

All data that support the plots within this paper and other findings of this study are available from the corresponding authors upon reasonable request.

## Acknowledgments

Z. L. and S.-F.S. acknowledge support from NYSTAR through Focus Center-NY–RPI Contract C180117. The device fabrication was supported by the Micro and Nanofabrication Clean Room (MNCR) at Rensselaer Polytechnic Institute (RPI). S.-F. S., D.X., and X.C. also acknowledge the support from NSF Grant DMR-1945420, DMR-2104902, and ECCS- 2139692. Y.-T.C. acknowledge support from NSF under award DMR-2104805. The optical spectroscopy measurements were supported by a DURIP award through Grant FA9550-20-1-0179. ST acknowledges support from NSF DMR-1904716, DMR-1838443, CMMI-1933214, and DOE-SC0020653. KW and TT acknowledge support from JSPS KAKENHI (Grant Numbers 19H05790, 20H00354, and 21H05233). YS and CZ acknowledge support from NSF PHY-2110212, PHY-1806227, ARO (W911NF17-1-0128), and AFOSR (FA9550-20-1-0220).

## Author contributions

S.-F.S. conceived the project. ZL, DC, and YM fabricated devices. ZL performed measurements. MB and ST grew the TMDC crystals. TT and KW grew the BN crystals. S.-F. S, Y.-T.C, ZL, and DC analyzed the data. S.-F. S supervised the project. S.-F. S. and Y.-T. C. wrote the manuscript with input from all authors.

## Competing interests

The authors declare no competing interest.

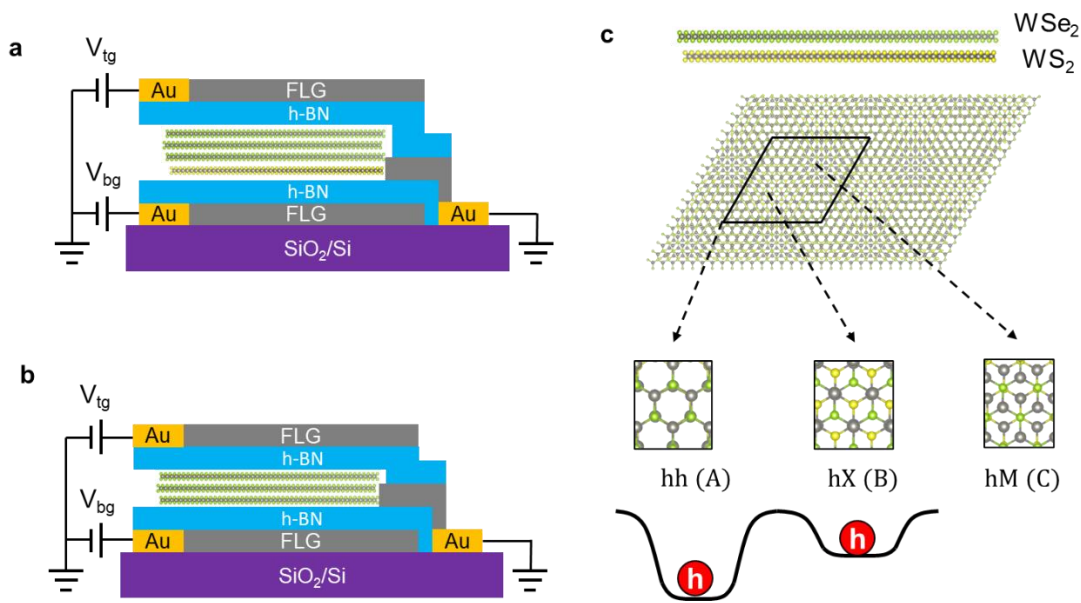

**Figure 1. Moiré site degree of freedom.** (a) and (b) are schematics of 3L $WSe_2$/ 1L $WS_2$ moiré heterojunction device and natural trilayer $WSe_2$ devices, respectively. Both devices are in a dual-gate configuration. (c) is the schematic of the $WSe_2/WS_2$ moiré superlattice, with three high symmetry points of C3 symmetry shown as hh, hX and hM. The naming convention of hh, hX and hM corresponds to aligning the hexagon center of the hole layer ($WSe_2$) with the hexagon center (h), the chalcogen atom (X), and the metal atom (M) of the electron layer ($WS_2$)[4], which we also call as A, B and C moiré sites for convenience. A and B sites are energy minima for holes and behave as quantum dots that confine carriers and excitons. We use the holes for illustration in (c), but the trapping of electrons and excitons will be similar.

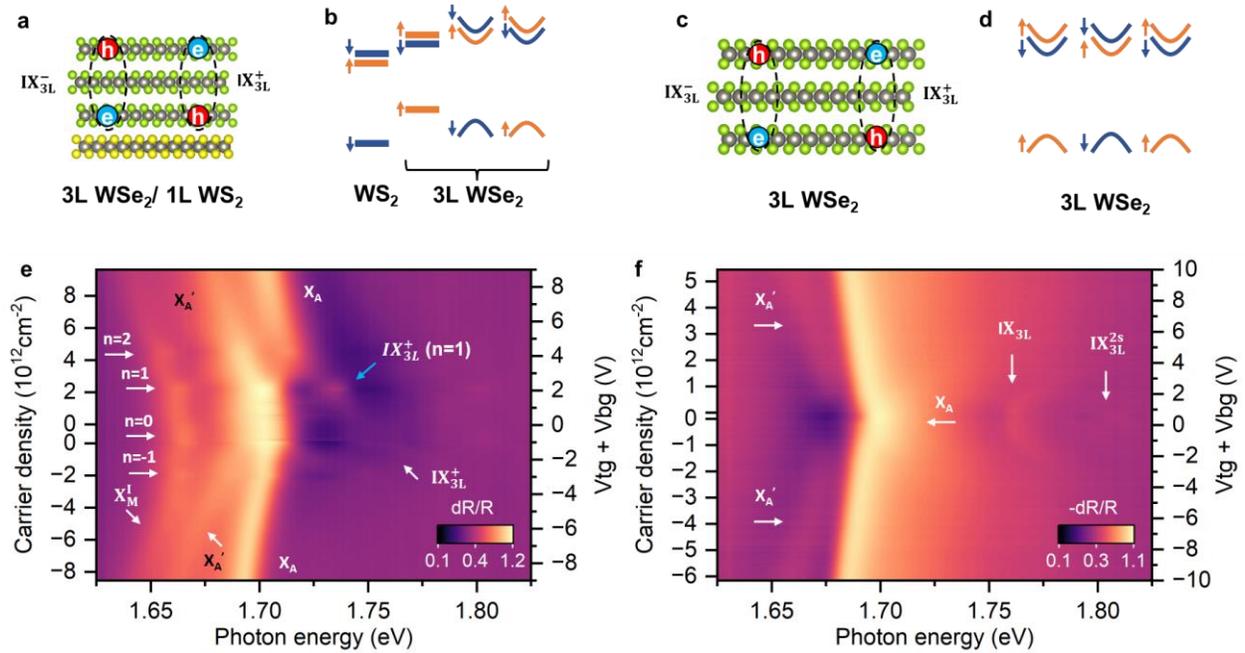

**Figure 2. Doping-dependent reflectance contrast spectra of 3L WSe₂/ 1L WS₂ and natural trilayer WSe₂ Devices.** (a), (b) and (e) are the schematic atomic structure, band alignment, and the doping-dependent reflectance contrast spectra of 3L WSe₂/ 1L WS₂ measured from device D1. (c), (d) and (f) are the schematic atomic structure, band alignment, and the doping-dependent reflectance contrast spectra of natural trilayer WSe₂ measured from device D2. The blue arrow in (e) denotes the enhanced reflection signal of the hybridized exciton (interlayer exciton $IX_{3L}^+$ hybridized with intrlayer exciton at the Mott insulator state at n=1.

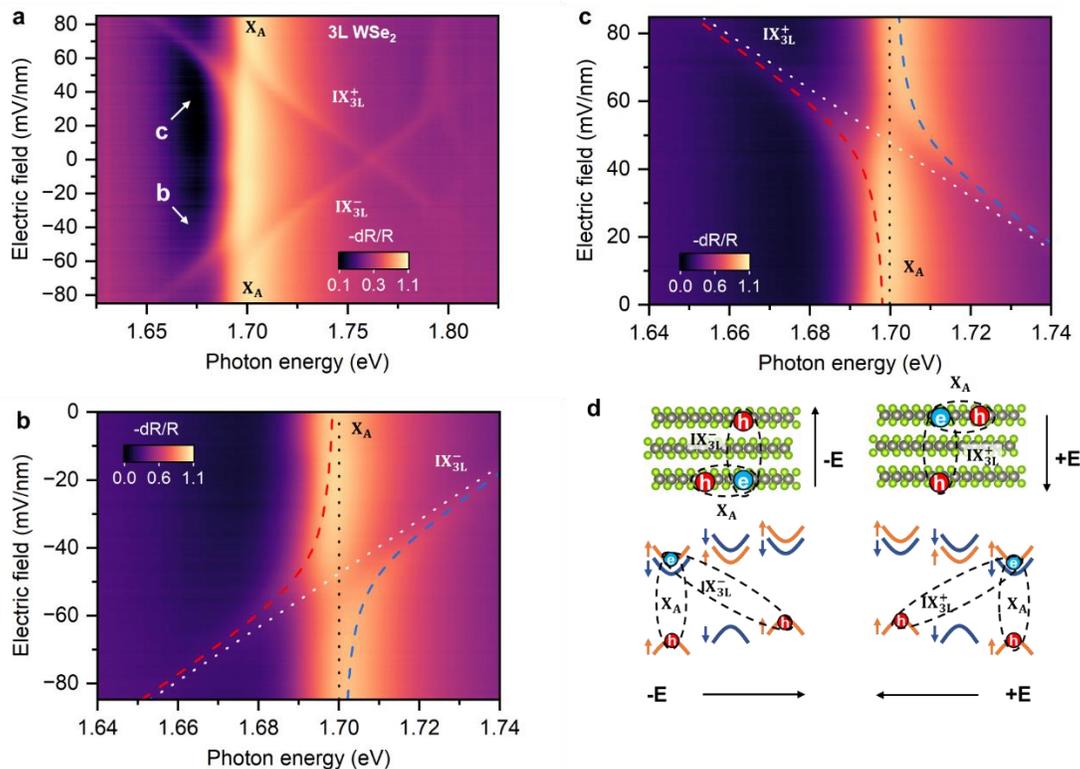

**Figure 3. Interlayer and intralayer exciton hybridization in natural trilayer WSe$_2$.** (a) shows the electric field dependence of reflectance contrast spectra measured from natural trilayer WSe$_2$ device (D2) plotted in log scale. (b) is the zoom-in of (a) in the region between 1.64 eV and 1.74 eV at negative electric fields plotted in linear scale. (c) is the zoom-in of (a) in the region between 1.64 eV and 1.74 eV at positive electric fields plotted in linear scale. The dashed red and blue lines show the fitting result of the hybridized excitonic states obtained by fitting the peak positions with a two-level hybridization model. The white and black dotted lines are the energies of unhybridized intralayer and interlayer excitons obtained from the fitting. (d) shows the schematics of the interlayer and intralayer excitons involved in the hybridization, both in real space and the band alignment configurations.

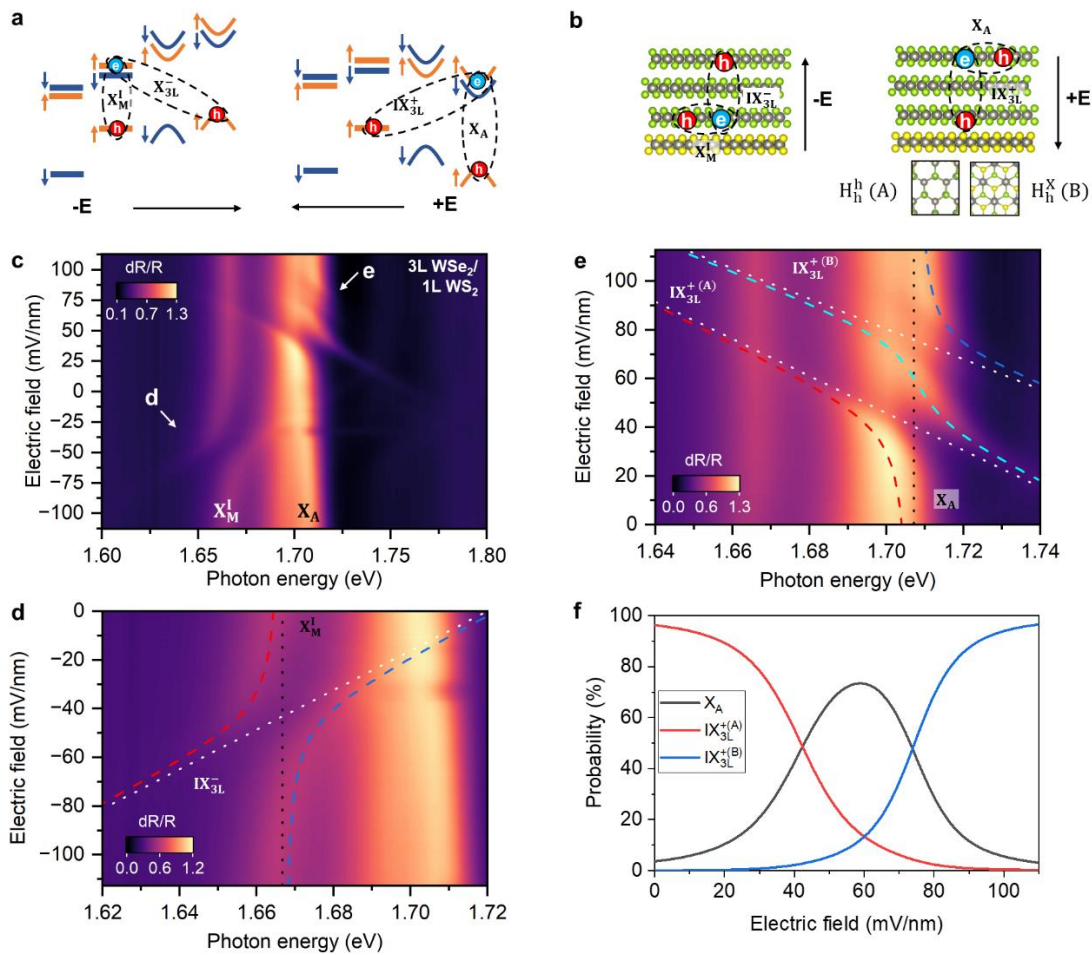

**Figure 4. Interlayer and intralayer exciton hybridization in the angle-aligned 3L WSe$_2$/ 1L WS$_2$ heterostructure.** (a) and (b) show the schematic band alignment and the real-space distribution of the hybridized excitons in 3L WSe$_2$/ 1L WS$_2$ moiré heterojunction. (c) shows the electric field dependence of reflectance contrast spectra measured from 3L WSe$_2$/ 1L WS$_2$ device (D1). (d) is the zoom-in of (a) in the region between 1.62 eV and 1.72 eV at negative electric fields. (e) is the zoom-in of (a) in the region between 1.64 eV and 1.74 eV at positive electric fields. The dashed lines show the hybridized excitonic states obtained by fitting the peak positions. The dotted lines are the energies of unhybridized intralayer and interlayer excitons obtained from the fitting. A two-level hybridization model using one intralayer exciton and one interlayer exciton as bases is used to fit the peak positions in (d), while a three-level hybridization model with one intralayer exciton and two different interlayer excitons is used to fit the peak positions in (e). (f) shows the fractional composition of the hybridized exciton corresponding to the cyan line as a function of the electric field, expressed as the probability of each interlayer or intralayer exciton.

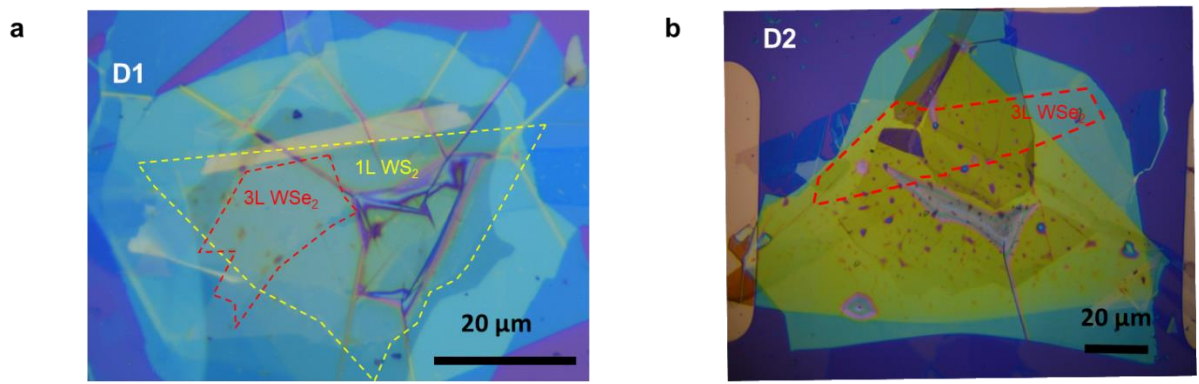

**Extended Fig. 1. Schematics and optical images of 3L WSe$_2$/ 1L WS$_2$ (D1) and 3L WSe$_2$ (D2) devices shown in Fig.1.** (a) and (b) show the schematic and the optical image of device D1 and D2, respectively.

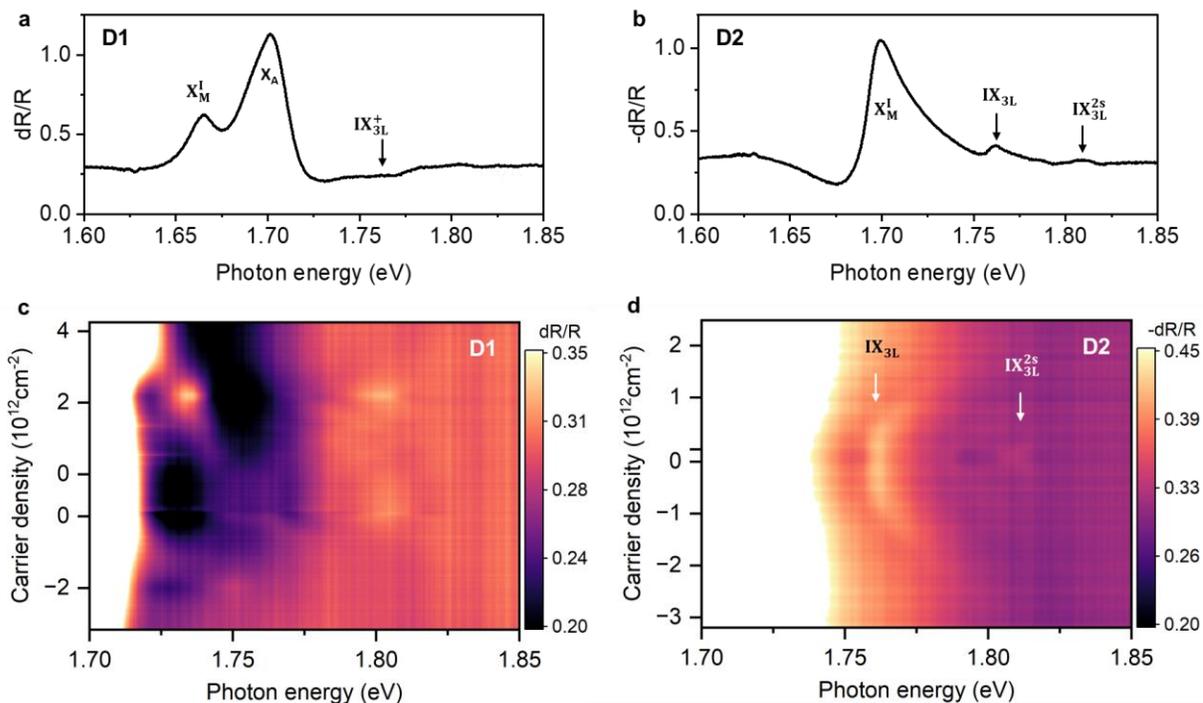

**Extended Fig. 2. Detailed reflectance contrast of device D1 and D2.** (a) and (b) are the reflectance contrast spectra of device D1 and D2 at charge neutral. (c) and (d) are the zoomed-in gate dependent reflectance contrast spectra of Fig. 2e (c) and Fig. 2f (d).

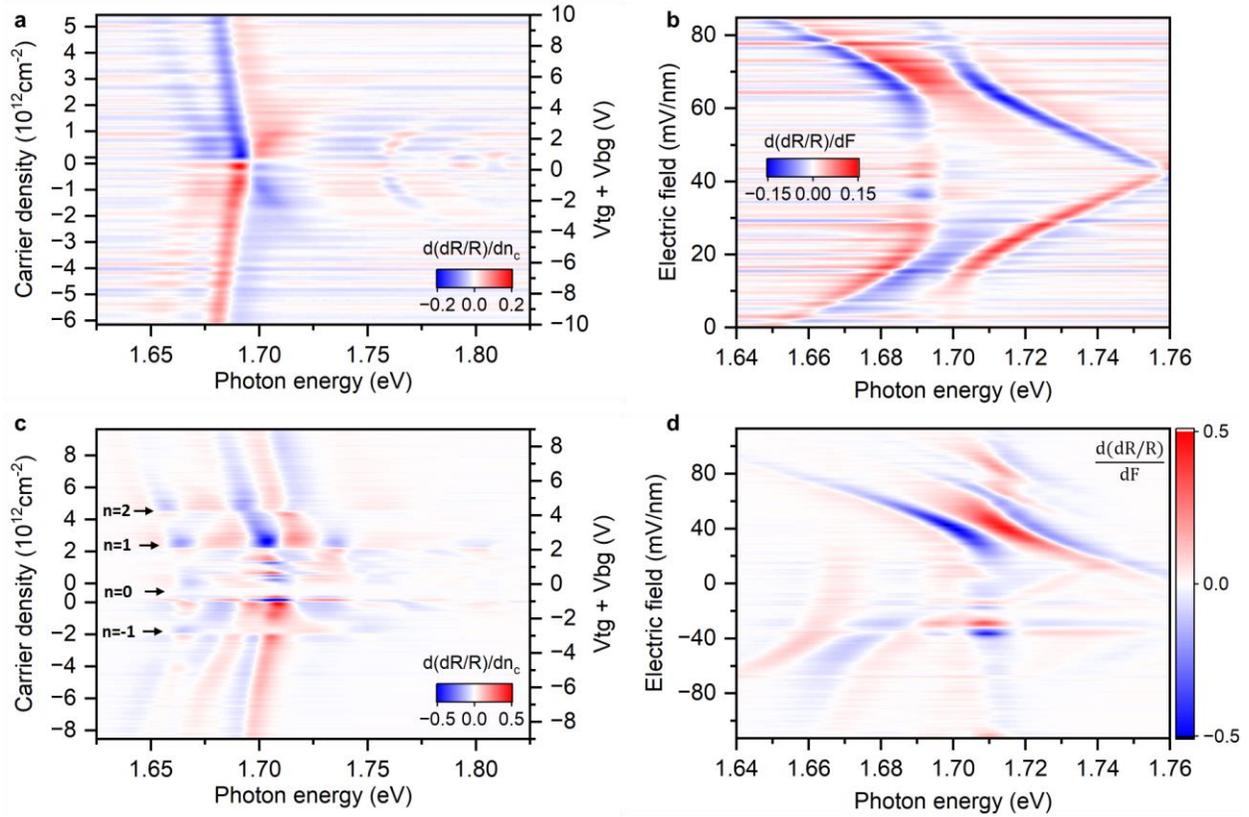

**Extended Fig. 3. Derivative analysis of the reflectance contrast spectra of device D1 and D2.** (a) and (c) show the derivative of dR/R from Fig. 2 f and e with respect to carrier density, respectively. (b) and (d) show the derivative of dR/R from Fig.3a and Fig.4c with respect to electric field, respectively.

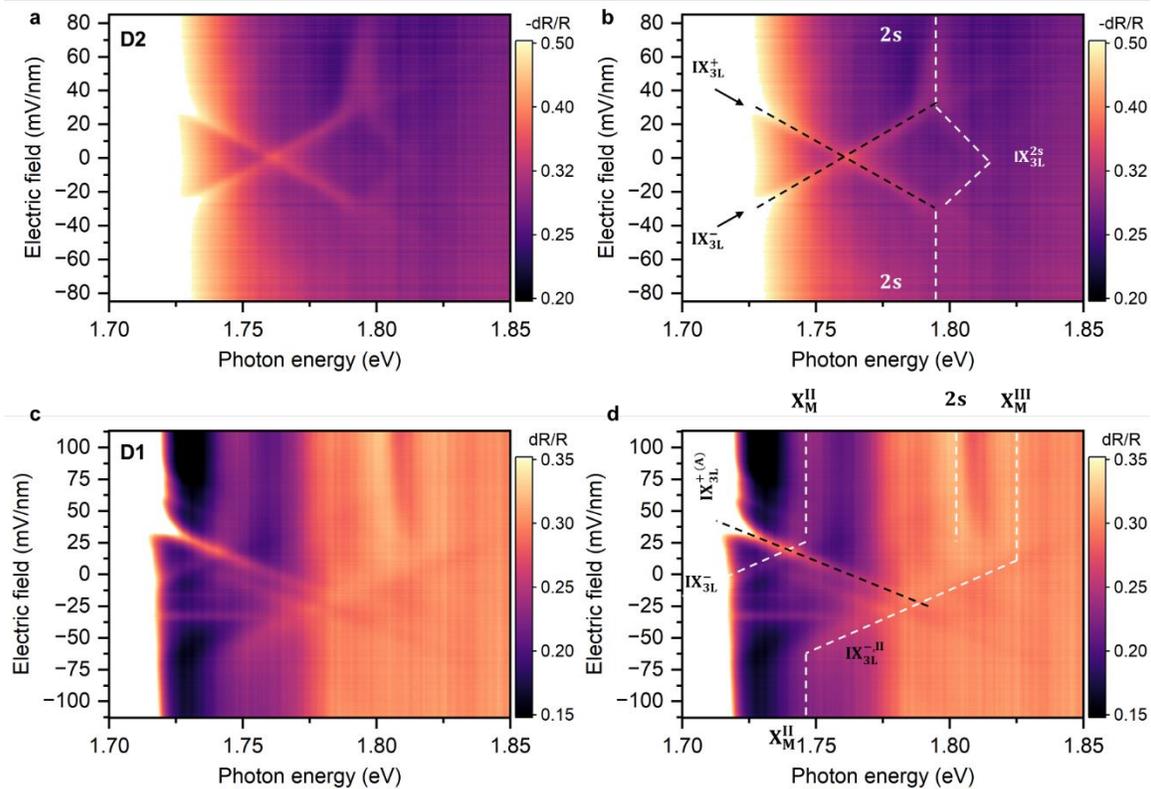

**Extended Fig. 4. Reflectance contrast spectra of zoom-in regions of Fig. 3a (a, b) and Fig. 4a (c, d).** (b) and (d) label the features in (a) and (c) with dashed lines as eye guides, respectively.

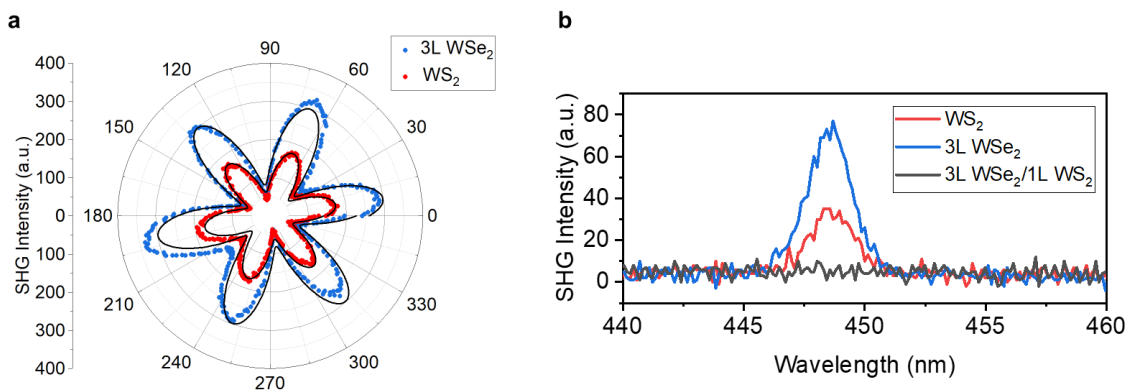

**Extended Fig. 5. Second-harmonic generation (SHG) spectra as a function of polarization angle from 3L WSe$_2$/ 1L WS$_2$ device D1.** (a) shows the integrated SHG intensity as a function of polarization angle from the 3L WSe$_2$ region and 1L WS$_2$ region. (b) shows the SHG spectra from 3L WSe$_2$, 1L WS$_2$ and 3L WSe$_2$/ 1L WS$_2$ heterostructure. The quench of the SHG signal on 3L WSe$_2$/ 1L WS$_2$ indicates the alignment angle is close to 60°. By fitting the polarization angle dependence of the SHG signal with a sinusoidal function, we determine the twist angle to be 0.9° ± 0.5°.

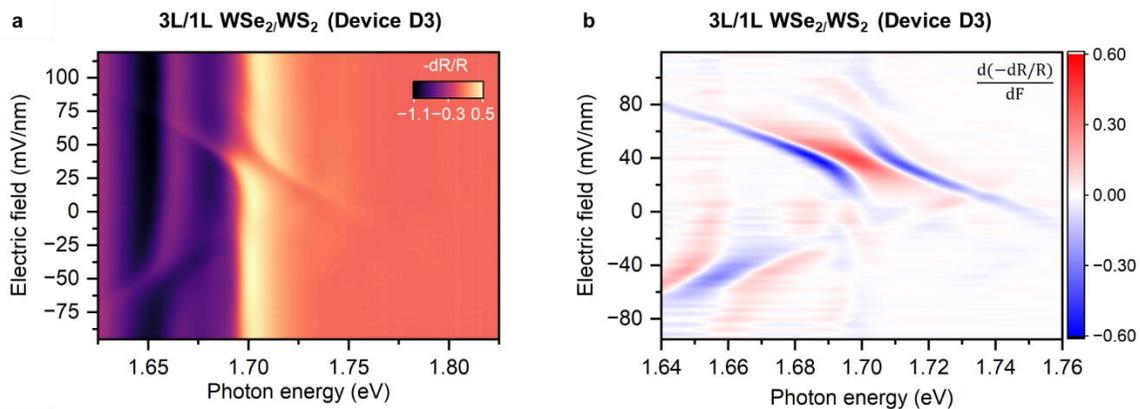

**Extended Fig. 6. Reflectance contrast spectra and the derivative of reflectance contrast with respect to the electric field of another 3L WSe$_2$/ 1L WS$_2$ device(D3).** (a) and (b) show dR/R and d(dR/R)/dF measured from 3L WSe$_2$/ 1L WS$_2$ device (D3).

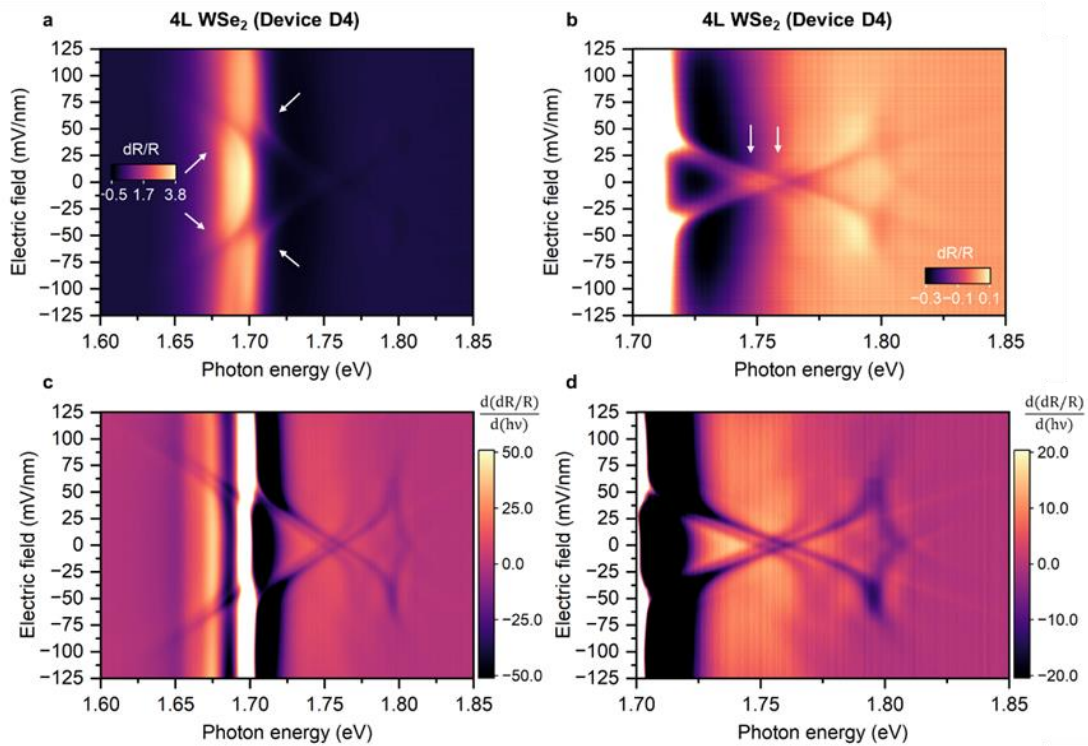

**Extended Fig. 7. Electric-field-dependent reflectance contrast spectra of dual gated natural 4L WSe$_2$ regions of device D4.** (a) and (b) are spectra from the 4L region. (c) and (d) are derivatives of reflectance contrast with respect to photon energy corresponding to (a) and (b), respectively.

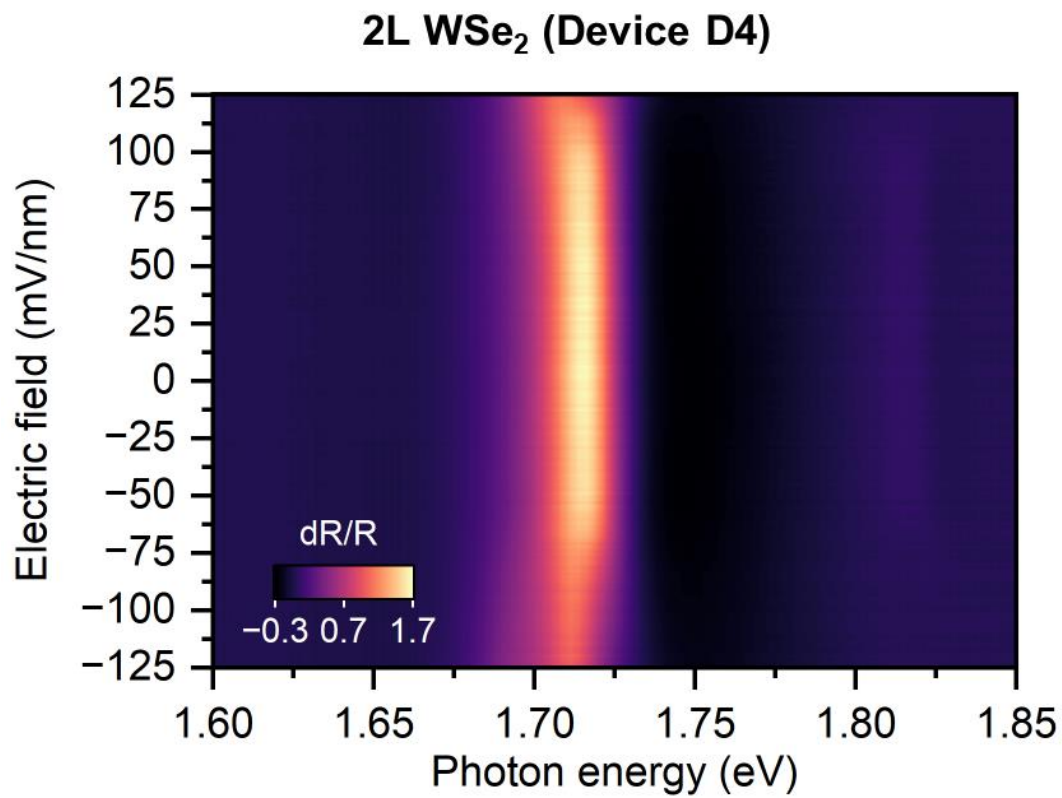

**Extended Fig. 8. Electric-field-dependent reflectance contrast spectra from a dual-gated 2H bilayer WSe$_2$ Device.**

Lian, Z., Chen, D., Meng, Y. *et al.* Exciton Superposition across Moiré States in a Semiconducting Moiré Superlattice. *Nat Commun* **14**, 5042 (2023).

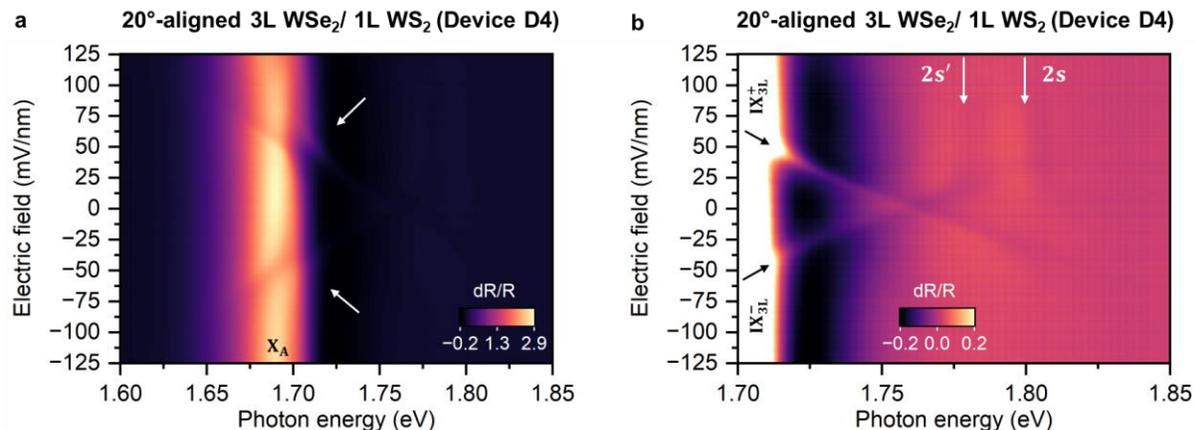

**Extended Fig. 9. Reflectance contrast spectra of a dual-gated 3L WSe$_2$/ 1L WS$_2$ device with a 20-degree twist angle.** (a) and (b) are the electric-field-dependent spectra for different photon energy ranges. The moiré exciton $X_M^I$ is absent in (a), and the hybridization of interlayer exciton $X_{3L}^i$ is symmetric about electric field in both (a) and (b).

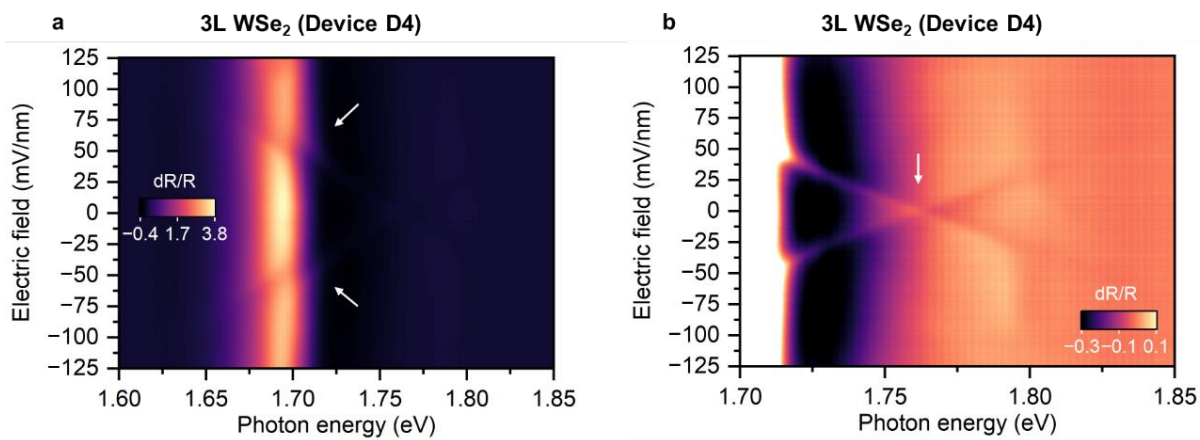

**Extended Fig. 10. Electric-field-dependent reflectance contrast spectra of dual gated natural 3L WSe$_2$ regions of device D4.** (a) and (b) are spectra from the 3L region.

# Supplementary Information

# Exciton Superposition across Moiré States in a Semiconducting Moiré Superlattice

Zhen Lian[1#], Dongxue Chen[1#], Yuze Meng[1], Xiaotong Chen[1], Ying Su[2], Rounak Banerjee[3], Takashi Taniguchi[4], Kenji Watanabe[5], Sefaattin Tongay[3], Chuanwei Zhang[2], Yong-Tao Cui[6*], Su-Fei Shi[1,7*]

1. Department of Chemical and Biological Engineering, Rensselaer Polytechnic Institute, Troy, NY 12180, USA
2. Department of Physics, University of Texas, Dallas, Texas, 75083, USA
3. School for Engineering of Matter, Transport and Energy, Arizona State University, Tempe, AZ 85287, USA
4. International Center for Materials Nanoarchitectonics, National Institute for Materials Science, 1-1 Namiki, Tsukuba 305-0044, Japan
5. Research Center for Functional Materials, National Institute for Materials Science, 1-1 Namiki, Tsukuba 305-0044, Japan
6. Department of Physics and Astronomy, University of California, Riverside, California, 92521, USA
7. Department of Electrical, Computer & Systems Engineering, Rensselaer Polytechnic Institute, Troy, NY 12180, USA

[#] These authors contributed equally to this work
[*] Corresponding authors: shis2@rpi.edu, yongtao.cui@ucr.edu

**Summary of contents**

Supplementary section 1: Fitting of the reflectance background

Supplementary section 2: Extracted peak positions from Fig.2a and Fig.3a

Supplementary section 3: Details of the modeling of hybridized excitons

Supplementary section 4: Discussion of the effect of doping on exciton energies

Supplementary section 5: Electric field dependence of reflectance contrast spectra at different filling factors measured from device D1

Supplementary section 6: Discussion of hole and electron hybridization scenarios

**Supplementary section 1: Fitting of the reflectance background**

We use the reflectance contrast spectrum $R_h$ at high hole-doping level ($> 6 \times 10^{12}\ cm^{-2}$) to construct the reflectance background for each spot measured. The high-energy features on the reflectance spectra, including moiré excitons and excited states, disappear due to the screening effect at such a doping level. The features corresponding to exciton resonance $X_M^I$, $X_A'$ and $X_A$ are identified by comparing $R_h$ with the reflectance spectra from regions with only h-BN and are then removed from the background. The rest of the spectrum is fitted by a polynomial function to construct a flat reflectance background $R_0$.

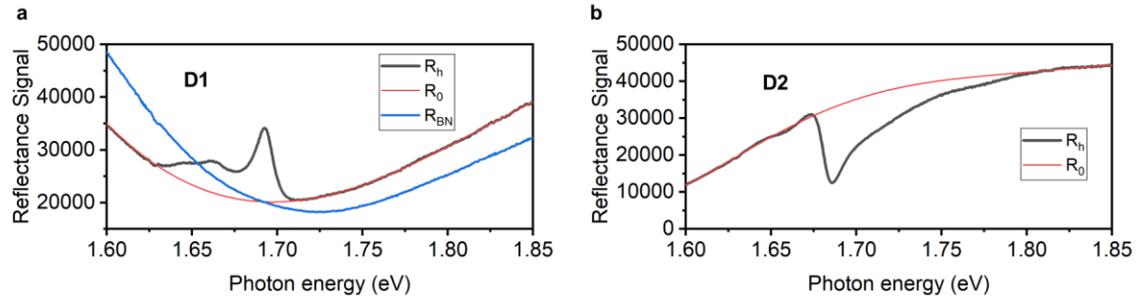

**Fig. S1 Fitting of the reflectance background of device D1 and device D2.** (a) shows the reflectance spectrum measured on 3L/1L WSe$_2$/WS$_2$ at a high doping level, the reflectance spectrum measured on h-BN, and the fitted reflectance background of device D1. (b) shows the reflectance spectrum measured on 3L WSe$_2$ and the fitted reflectance background of device D2.

**Supplementary section 2: Extracted peak positions from Fig.3a and Fig.4c**

Fig. S2 shows the extracted peak positions from Fig.3a and Fig.4c of the main text, which is further used to fit the hybridized exciton models described in supplementary section 3.

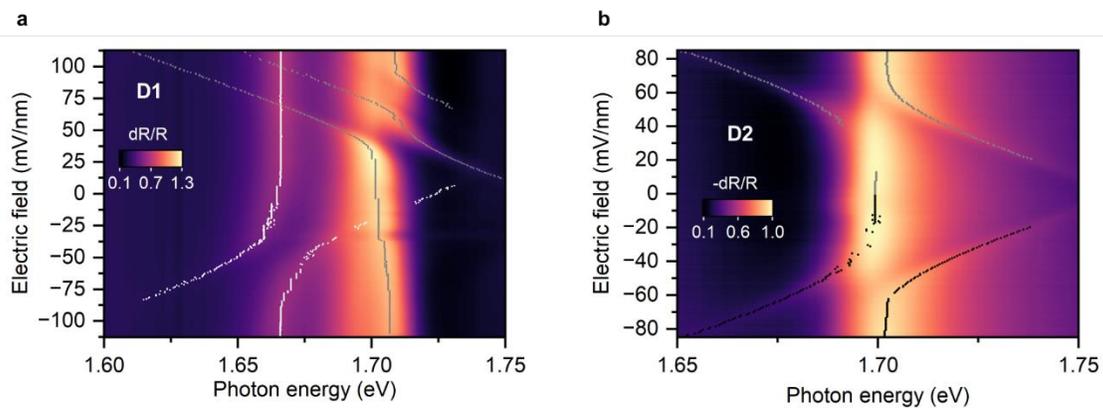

**Fig. S2. Extracted peak positions from the electric field dependence of reflectance contrast spectra of devices D1 and D2.** (a) and (b) are the extracted peak positions overlayed on the colorplot of Fig.3a and Fig.4c. The dots indicate the fitted peaks.

**Supplementary section 3: Details of the Modeling of hybridized excitons**

We model the intralayer exciton energy as a constant under a vertical electric field, which is denoted as Xa. The interlayer exciton energy can be expressed as $Xi(F) = Xi0 - e \cdot d \cdot F$, where Xi0 is the interlayer exciton energy at 0 electric field, e is the elementary charge, d is the interlayer distance (defined as positive when the dipole moment is downward and negative when upward) and F is the vertical electric field inside the heterostructure. In Fig. 3b, 3c, and 4d, the hybridized exciton can be modeled by considering the hybridization between one intralayer exciton and one interlayer exciton, which is given by the following Hamiltonian:

$$\begin{bmatrix} Xa & \Delta 1 \\ \Delta 1 & Xi1 - e \cdot d1 \cdot F \end{bmatrix}$$

where Δ is the strength of hybridization, and it is a constant in our model.

In Fig. 4e, there are two interlayer excitons involved in the hybridization, and the Hamiltonian is given by:

$$\begin{bmatrix} Xa & \Delta 1 & \Delta 2 \\ \Delta 1 & Xi1 - e \cdot d1 \cdot F & 0 \\ \Delta 2 & 0 & Xi2 - e \cdot d2 \cdot F \end{bmatrix}$$

The eigenvalues $\xi_j$ of the Hamiltonian give the energies of the hybridized exciton branches.

The extracted peak positions are fitted using F as an independent variable, $\xi_j$ as a dependent variable, and $Xa$, $Xi0$, $\Delta$ and d as unknown parameters. A gradient search approach is used to minimize the mean squared error, defined as $MSE = \frac{1}{N} \sum_F \sum_j (y_j(F) - \xi_j(F))^2$. The fitting results from devices D1 and D2 are listed in Table S1.

| Device | D2 | | D1 | |
|---|---|---|---|---|
| Hybridized exciton | Xh(-) | Xh(+) | Xh(-) | Xh(+) |
| Xa (eV) | 1.7001±0.0001 | 1.6999±0.0001 | 1.6668±0.0001 | 1.7071±0.0001 |
| Xi1 (eV) | 1.7601±0.0001 | 1.7603±0.0003 | 1.7200±0.0001 | 1.7601±0.0002 |
| Xi2 (eV) | -- | -- | -- | 1.8293±0.0003 |
| d1 (nm) | -1.262±0.003 | 1.262±0.008 | -1.231±0.003 | 1.313±0.004 |
| d2 (nm) | -- | -- | -- | 1.609±0.004 |
| Δ1 (meV) | 10.1±0.1 | 10.7±0.3 | 11.4±0.1 | 11.1±0.2 |
| Δ2 (meV) | -- | -- | -- | 10.9±0.2 |

**Table S1. Summary of the hybridized exciton model parameters extracted from fitting.**

**Supplementary section 4: Discussion of the effect of doping on exciton energies**

Adopting the dielectric constants of 3.5 for hBN, 6.3 for monolayer $WS_2$, and 7.5 for monolayer $WSe_2$[1], in the natural 3L $WSe_2$ device, the second $WSe_2$ layer has a higher average permittivity environment comparing to the third $WSe_2$ layer considering the configuration of different $WSe_2$ layers. The higher dielectric constant can tune more bandgap and binding energy and will induce a corresponding larger energy redshift of intralayer exciton[2,3,] so we attributed the two branches labeled as $X_A$ and $X_A'$ in Fig. 2f to the intralayer exciton in the first/third and second layer $WSe_2$, respectively.

As both $X_A$ and $X_A'$ redshift linearly with increasing doping under no matter hole or electron doping, we calculated the slopes of doping-dependent energy shifts shown in Table S2, which represent the sensitivity of the intralayer exciton energy to the doping level. The slopes at n doping and p doping for $X_A$ and $X_A'$ are similar, and the slope of $X_A'$ is larger than $X_A$'s. In addition, the dielectric constant of the doped layers will increase because of the accumulation of free carriers. Correspondingly, this overall increase of permittivity environment will lead to a doping-dependent redshift of both $X_A'$ and $X_A$. However, due to the configuration difference between the second and third $WSe_2$ layer; that is both doped first/third $WSe_2$ layers adjacent to the second $WSe_2$ layer but the two doped first/second (second/third) $WSe_2$ layers on the bottom (top) side of the third (first) $WSe_2$ layer, doping will induce a more significant average dielectric environment modification for the second $WSe_2$ layer than the first/third $WSe_2$ layer which accounts for a more sensitive doping dependent energy redshift of the intralayer exciton in second $WSe_2$ layer.

Analogous to the discussion of natural trilayer $WSe_2$, for the 3L/1L device, we conclude that $X_A$ originates from the third layer $WSe_2$ and $X_A'$ from the second layer $WSe_2$. The same local dielectric screening mechanism as natural 3L $WSe_2$ results in the doping-dependent redshift of $X_A$ and $X_A'$ at the n doping region. At the p doping region, the larger doping sensitivity can be explained by the fact that the second $WSe_2$ layer is closer to n-doped $WS_2$ than the third $WSe_2$. In addition, the energy jump of $X_A$ between doping levels n=1 and n=2 results from the doping dependent hybridization between interlayer exciton and intralayer exciton, which is discussed in Fig. S3.

| 3L | $X_A'$ (p doping) | $X_A$ (p doping) | $X_A'$ (n doping) | $X_A$ (n doping) |
|---|---|---|---|---|
| Energy shift/doping (unit meV/ $10^{12} cm^{-2}$) | 2.80 | 1.26 | 2.66 | 1.35 |
| 3L/1L | $X_A'$ (p doping) | $X_A$ (p doping) | $X_A'$ (n doping) | $X_A$ (n doping) |
| Energy shift/doping (unit meV/ $10^{12} cm^{-2}$) | 2.06 | 1.01 | 2.05 | 1.04 |

**Table S2. Summary of the energy shifts of $X_A'$ and $X_A$ as functions of carrier density.**

It is worth noting that although the slopes for $X_A'$ and $X_A$ are different for the natural trilayer (3L) $WSe_2$ device and the 3L/1L $WSe_2/WS_2$ moiré heterojunction device, likely due to dielectric environment difference, the ratio of the slope of $X_A'$ to that of $X_A$ is about 2.0, similar to that of natural trilayer $WSe_2$ (2.2 for p-doping and 2.0 for n-doping). This further confirms that the nature of $X_A'$ and $X_A$ observed in 3L/1L $WSe_2/WS_2$ moiré heterojunction is the same as that of natural trilayer $WSe_2$.

**Supplementary section 5: Electric field dependence of reflectance contrast spectra at different filling factors measured from device D1**

For different filling factor n (number of electrons or holes per moiré supercell, "+" for electrons and "-" for holes), we studied the electric field dependent reflectance contrast spectra of 3L/1L $WSe_2/WS_2$ device (D1, shown in the main text) in Fig.S3. It is evident that as the electron doping increases, the hybridization electric field for the two interlayer excitons and intralayer exciton shifts to a smaller magnitude, approaching zero for n=2. This is because the electrostatically introduced electrons go to the $WS_2$ layer due to the type II alignment. As a result, the $WSe_2$ layers will experience an effective electric field pointing towards the $WS_2$ layer same direction as the positive electric field needed to realize the hybridization of the interlayer moiré excitons with intralayer excitons in the third $WSe_2$ layer.

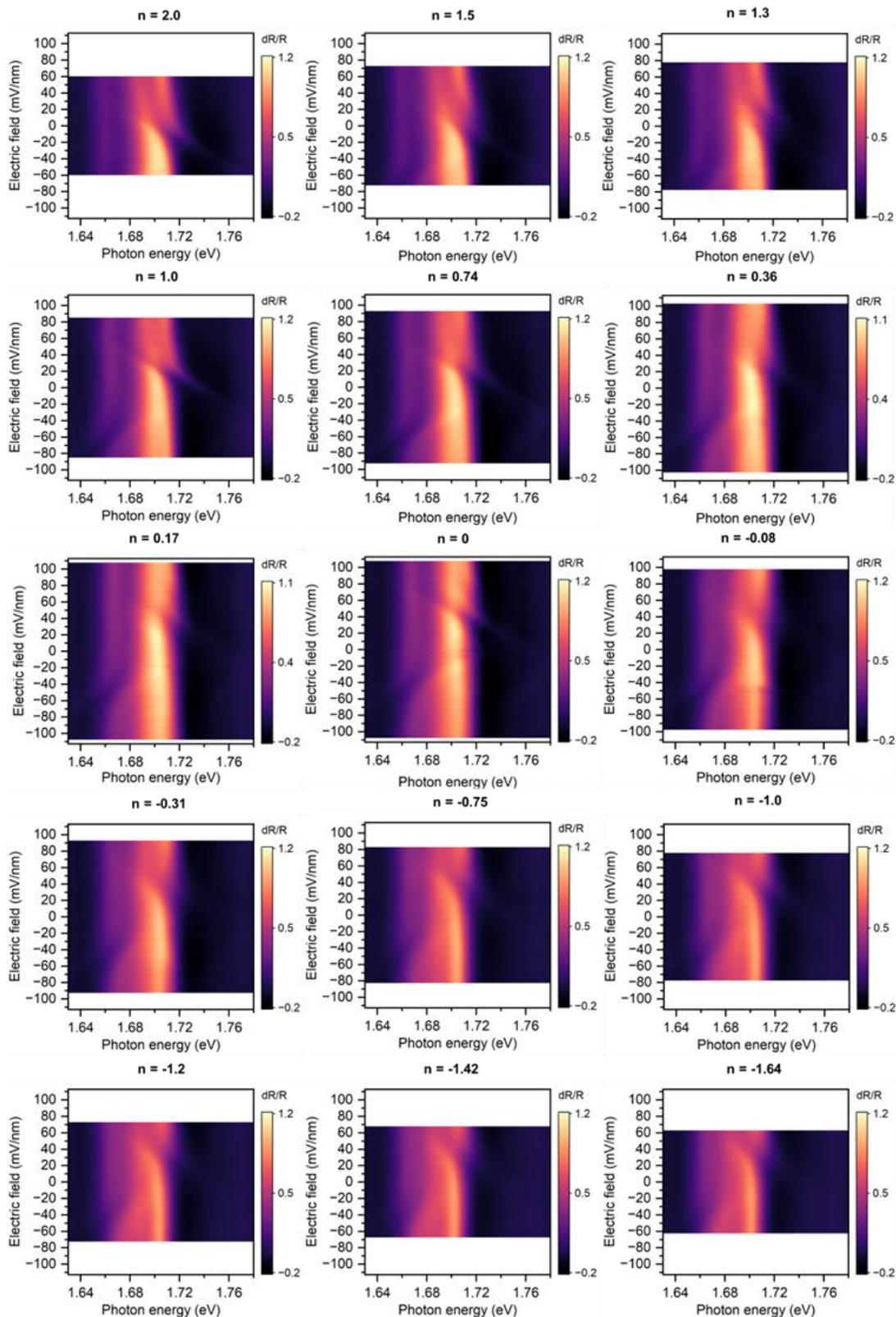

**Fig.S3. Electric field dependence of reflectance contrast spectra at different filling factors measured from device D1.**

**Supplementary section 6: Discussion of hole and electron hybridization scenarios**

Considering the up and down interlayer exciton dipoles, along with the possibilities of electron or hole tunneling, the different hybridization scenarios are schematically illustrated in Fig. S4a, b, d, e. Fig. S4b and Fig. S4d correspond to the scenarios of hole tunneling, which have been discussed in the main text. Fig. S4a and Fig. S4e correspond to the hybridization of $IX_{3L}^+$ with $X_M^I$, and the hybridization of $IX_{3L}^-$ with $X_A$ when electron tunneling is considered. It is clear from Fig. S4c that associated excitons cross each other without level avoiding (white arrows marked by 'a' and 'e'), confirming the validity of our discussion in the main text that ignore the conduction band hybridization.

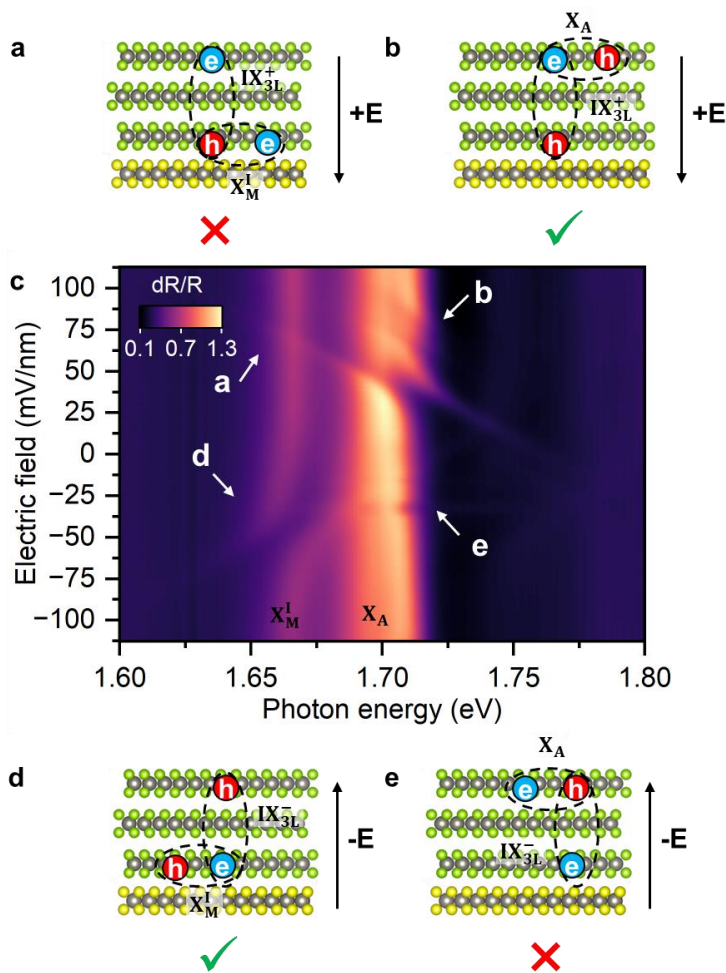

**Fig.S4. Different hybridization scenarios in 3L/1L WSe$_2$/WS$_2$.** (a) and (b) are the schematics showing the hybridization of intralayer exciton with interlayer exciton $IX_{3L}^+$. (d) and (e) are the schematics showing the hybridization of interlayer excitons with interlayer exciton $IX_{3L}^-$. (c) shows the same data as in Fig. 4c, with white arrows indicating the positions where hybridization can possibly occur, considering both electron and hole tunneling.

Lian, Z., Chen, D., Meng, Y. *et al.* Exciton Superposition across Moiré States in a Semiconducting Moiré Superlattice. *Nat Commun* **14**, 5042 (2023).